\documentclass[conference,a4paper]{IEEEtran}

\usepackage{amsmath,amssymb,mathrsfs,amsbsy}
\usepackage{cite}
\usepackage{subfigure}
\usepackage{graphicx,color}
\usepackage{hyperref}
\usepackage{subfigure}
\usepackage{balance}
\usepackage{tikz,siunitx}
\usepackage{url}

\newtheorem{lem}{Lemma}

\newtheorem{thm}{Theorem}

\newtheorem{rmk}{Remark}

\DeclareMathOperator*{\argmin}{arg\,min}

\definecolor{sblue}{RGB}{0,51,120}
\definecolor{sred}{RGB}{200,51,130}

\ifCLASSINFOpdf
\else
\fi

\begin{document}

\title{\Huge Sherman-Morrison Regularization for ELAA Iterative Linear Precoding}

\author{Jinfei Wang, Yi Ma, Na Yi, and Rahim Tafazolli\\
	{\small 5GIC and 6GIC, Institute for Communication Systems, University of Surrey, Guildford, UK, GU2 7XH}\\
		{\small Emails: (jinfei.wang, y.ma, n.yi, r.tafazolli)@surrey.ac.uk}\\
		}
\markboth{}%
{}

\maketitle

\begin{abstract}
The design of iterative linear precoding is recently challenged by extremely large aperture array (ELAA) systems, where conventional preconditioning techniques could hardly improve the channel condition.
In this paper, it is proposed to regularize the extreme singular values to improve the channel condition by deducting a rank-one matrix from the Wishart matrix of the channel.
Our analysis proves the feasibility to reduce the largest singular value or to increase multiple small singular values with a rank-one matrix when the singular value decomposition of the channel is available.
Knowing the feasibility, we propose a low-complexity approach where an approximation of the regularization matrix can be obtained based on the statistical property of the channel.
It is demonstrated, through simulation results, that the proposed low-complexity approach significantly outperforms current preconditioning techniques in terms of reduced iteration number { for more than $10\%$} in both ELAA systems as well as symmetric multi-antenna (i.e., MIMO) systems when the channel is i.i.d. Rayleigh fading.	
\end{abstract}

\section{Introduction}\label{secI}
The iterative linear precoding can provide the inverse of matrices with square-level complexity.
This is important for real-time signal processing, since the complexity of exact matrix inverse is at cubic-level and is beyond current hardware capability \cite{Kammoun2014}.
However, the ill-condition of the channel will significantly increase the iteration number needed to converge to the exact linear precoder.
This issue is particularly important in the recent studies of extremely large aperture array (ELAA) systems, where the users fall in the near field of the service-antenna array.
In this case, the wireless channel becomes very ill-conditioned due to the spatial non-stationarity, which is caused by various factors such as large-scale fading, shadowing and mixed line-of-sight (LoS)/non-LoS antenna links \cite{9170651,9685536}.

Improving the channel condition to accelerate the convergence of the iterative linear precoding has been widely studied in the recent decades for conventional massive multiple-input multiple-output (MIMO) systems.
This is usually realized by multiplying a preconditioning matrix to the Wishart matrix of the channel, such as the Jacobi \cite{8417575}, Gauss-Seidel (GS) \cite{Lee2020} and symmetric successive over-relaxation (SSOR) preconditioning \cite{7399337}.
However, these approaches are mainly designed to inverse diagonal-dominant matrices (i.e., the antenna links are nearly orthogonal to each other).
In ELAA system, this assumption can no longer be suitable due to the existence of LoS links as well as the large number of user antennas.
These discussions motivate our investigation to improve the iterative linear precoding in ELAA systems.

In this paper, we propose to regularize the extreme singular values of the Wishart matrix of the channel with a rank-one matrix.
To this end, we first analyze the feasibility to reduce the largest singular value (see \textit{Theorem \ref{thm1}}) or to increase multiple small singular values (see \textit{Theorem \ref{thm2}}) with a rank-one matrix when the singular value decomposition (SVD) of the channel is available.
With the feasibility analysis in mind, we also propose a low-complexity approach that approximates the rank-one regularization matrix based on the difference between the Wishart matrix of the channel and a scaled identity matrix.
This approach is general for different MIMO systems as long as the scale of the identity matrix needs to be adapted based on the statistical property of the channel, such as the LoS-dominated ELAA (see \textit{Remark \ref{rmk1}}) and symmetric MIMO (see \textit{Remark \ref{rmk2}}).
Moreover, the proposed approach can be combined with the list algorithm to mitigate the sub-optimality of the approximation.

{ In the simulations, the proposed approach is compared to the iterative linear precoding with or without preconditioning, where the preconditioning techniques include} the Jacobi, GS, SSOR preconditioning.
In the LoS-dominated ELAA system, the proposed approach can reduce the iteration number for more than $40\%$, while the preconditioning techniques can hardly improve the performance of the HB algorithm.
In the symmetric MIMO when channel is i.i.d. Rayleigh, the proposed approach can reduce the iteration number for around $7\%$.
And with the help of list algorithm, this improvement increases to around $20\%$.
While in a symmetric ELAA system, the proposed approach reduces the iteration number for around $4\%$.
With the help of list algorithm, this improvement increases to around $12\%$.
Meanwhile, current preconditioning techniques could still hardly improve the iteration number.

\section{System Model and Problem Statement}
Consider a MIMO downlink system with $M$ transmit-antennas and $N$ receive-antennas ($M\geq N$), where $N$ independent data-streams are transmitted to the users.
Denote the wireless channel by $\mathbf{H}\in\mathbb{C}^{N\times M}$.
To avoid interference across data-streams, a linear precoder $\mathbf{W}\in\mathbb{C}^{M\times N}$ is employed for interference rejection:
\begin{equation}
\mathbf{y}=P_\mathrm{t}\mathbf{H}\mathbf{W}\mathbf{s}+\mathbf{v},
\end{equation}
where $\mathbf{y}\in\mathbb{C}^{N\times1}$ stands for the received signals, $P_\mathrm{t}$ for the transmission power, $\mathbf{s}\in\mathbb{C}^{N\times1}$ for the information-bearing symbols, $\mathbf{v}\in\mathbb{C}^{N\times1}$ ($\mathbf{v}\sim\mathcal{CN}(\mathbf{0},N_0\mathbf{I})$) for the additive white Gaussian noise (AWGN), $\mathbf{I}$ for the identity matrix.

To cancel the interference, the linear precoder $\mathbf{W}$ is usually designed to be the pseudo inverse of the channel $\mathbf{H}$ (e.g., ZF and LMMSE \cite{1468466}).
This generally involves the matrix inverse of the Wishart matrix of $\mathbf{H}$ (denoted by {$\mathbf{A}\triangleq\mathbf{H}\mathbf{H}^H\in\mathbb{C}^{N\times N}$}):
\begin{equation}
\mathbf{W}=\mathbf{H}^H(\mathbf{H}\mathbf{H}^H)^{-1}=\mathbf{H}^H\mathbf{A}^{-1}.
\end{equation}
The computing process to obtain $\mathbf{A}^{-1}$ has a complexity of $\mathcal{O}(N^3)$ and cannot be parallelized \cite{Kammoun2014}.
This means obtaining the linear precoder $\mathbf{W}$ is beyond current hardware computation capability.

\subsection{Iterative Linear Precoding}
The idea of iterative linear precoding is to replace the calculation of $\mathbf{A}^{-1}$ by an iterative process, where the $t^\mathrm{th}$ iteration (${t=0,1,...}$) outputs a square matrix $\mathbf{X}_t\in\mathbb{C}^{N\times N}$ and $\lim\limits_{t\to\infty}\mathbf{X}_t=\mathbf{A}^{-1}$.

There are various ways to perform this iterative process, such as the Neumann series, the Richardson iteration and the Hotelling-Bodewig (HB) algorithm. 
{In this paper, the HB algorithm is employed for analysis and simulations in appreciation of its faster convergence speed (i.e., local quadratic).}
Specifically, the iteration of HB algorithm is given by \cite{Hotelling1943}
\begin{equation}\label{eq04}
\mathbf{X}_{t}=\mathbf{X}_{t-1} - 2\mathbf{X}_{t-1}\mathbf{A}\mathbf{X}_{t-1},~\mathbf{X}_0=\omega\mathbf{A}^H,
\end{equation}
where $\omega$ stands for a scaling factor to ensure that $\rho(\mathbf{I}-\omega\mathbf{A}\mathbf{A}^H)<1$, $\rho(\cdot)$ for the spectrum radius of a matrix.

\subsection{Convergence of Iterative Matrix Inverse}
The iterative process in \eqref{eq04} can be simplified as follows:
\begin{equation}\label{eq05}
\mathbf{I}-\mathbf{A}\mathbf{X}_{t}{=(\mathbf{I}-\mathbf{A}\mathbf{X}_0)^{2^t}}=(\mathbf{I}-\omega\mathbf{A}\mathbf{A}^H)^{2^t}.
\end{equation}
Since $\mathbf{A}$ is a Hermitian matrix, we have the SVD of $\mathbf{A}$ as $\mathbf{U}\mathbf{\Lambda}\mathbf{U}^H$, where $\mathbf{\Lambda}$ is a diagonal matrix with its elements $\lambda_n$, $_{n=0,...,N-1}$ standing for the singular values and $\lambda_{0}\geq...\geq\lambda_{N-1}>0$.
Then, \eqref{eq05} can be simplified as
\begin{equation}\label{eq06}
\mathbf{I}-\mathbf{A}\mathbf{X}_{t}=\mathbf{U}(\mathbf{I}-\omega\mathbf{\Lambda}^2)^{2^t}\mathbf{U}^H.
\end{equation}
Since $\rho(\mathbf{I}-\omega\mathbf{A}\mathbf{A}^H)<1$, \eqref{eq06} shows that $\mathbf{I}-\mathbf{A}\mathbf{X}_{t}$ (i.e., $\mathbf{X}_{t}$ tends to $\mathbf{A}^{-1}$) tends to zero as $t$ tends to infinity.
In practice, $\|\mathbf{I}-\mathbf{A}\mathbf{X}_{t}\|_\mathrm{F}^2$ will be negligible when $t$ is large enough, where $\|\cdot\|$ stands for the Frobenius norm.

However, when $t$ is fixed, $\|\mathbf{I}-\mathbf{A}\mathbf{X}_{t}\|_\mathrm{F}^2$ can increase with the decrease of $\omega$ and/or the increase of the condition number of $\mathbf{A}$ (denoted by $\kappa(\mathbf{A})$, $\kappa(\mathbf{A})\triangleq(\lambda_{0})/(\lambda_{N-1})$).
Specifically, we have $\|\mathbf{I}-\mathbf{A}\mathbf{X}_{t}\|_\mathrm{F}^2=\sum_{n=0}^{N-1}(1-\omega\lambda_{n})^{2^t}$.
In this paper, we employ the optimal $\omega$ (i.e., $\omega^\star$) for theoretical analysis of $\kappa(\mathbf{A})$: $\omega^\star$ is given by $\omega^\star{=}(2)/(\lambda_0^2+\lambda_{N-1}^2)$ \cite{Saad2003}.

Since $\lambda_{0}\geq...\geq\lambda_{N-1}$, the dominant terms in $\|\mathbf{I}-\mathbf{A}\mathbf{X}_{t}\|_\mathrm{F}^2$ are the first and last term of the summation. These two terms are equal and can be simplified as
\begin{equation}
\left|\frac{\lambda_{0}^2+\lambda_{N-1}^2-2\lambda_0^2}{\lambda_{0}^2+\lambda_{N-1}^2}\right|=\left|\frac{\lambda_{0}^2+\lambda_{N-1}^2-2\lambda_{N-1}^2}{\lambda_{0}^2+\lambda_{N-1}^2}\right|=\frac{\kappa(\mathbf{A})^2-1}{\kappa(\mathbf{A})^2+1}.
\end{equation} 
Therefore, by keeping the two dominant terms,  $\|\mathbf{I}-\mathbf{A}\mathbf{X}_{t}\|_\mathrm{F}^2$ can be approximated as
\begin{equation}
\|\mathbf{I}-\mathbf{A}\mathbf{X}_{t}\|_\mathrm{F}^2\approx2\left(\frac{\kappa(\mathbf{A})^2-1}{\kappa(\mathbf{A})^2+1}\right)^{2^{t}}.
\end{equation}
Since we have
\begin{equation}
\left(\partial \frac{\kappa(\mathbf{A})^2-1}{\kappa(\mathbf{A})^2+1}\right)/\left(\partial \kappa(\mathbf{A})\right)=\frac{4\kappa(\mathbf{A})}{(1+\kappa(\mathbf{A})^2)^2}>0,
\end{equation}
it is known that $\|\mathbf{X}_t-\mathbf{A}\|_\mathrm{F}^2$ increases with the increase of $\kappa(\mathbf{A})$.

\subsection{Prior Arts: Matrix Preconditioning}
Improving $\kappa(\mathbf{A})$ has been widely studied in the past decade (e.g., \cite{8417575,Lee2020,7399337}).
The conventional approach is to multiply $\mathbf{A}$ with a preconditioning matrix $\mathbf{P}\in\mathbb{C}^{N\times N}$ such that $\kappa(\mathbf{P}\mathbf{A})<\kappa(\mathbf{A})$.
Then, the iterative process outputs $\mathbf{X}_{t}\to(\mathbf{P}\mathbf{A})^{-1}=\mathbf{A}^{-1}\mathbf{P}^{-1}$ with a small iteration number, and $\mathbf{A}^{-1}$ is therefore known by $\mathbf{A}^{-1}=\mathbf{X}_t\mathbf{P}$.
However, these preconditioning approaches are mainly designed for diagonal-dominant matrices \cite{Saad2003,James1975}, which is the case of conventional massive-MIMO systems.

In ELAA systems, $\mathbf{A}$ can be no longer diagonal-dominant due to the spatial non-stationarity (e.g., the correlation of LoS antenna links).
Moreover, compared to conventional massive-MIMO ($\kappa(\mathbf{A})$ is usually below $10$), the channel is very ill-conditioned ($\kappa(\mathbf{A})$ easily increases to $10^2$ or more) in ELAA systems.
These discussions show that handling the condition number is a major problem for iterative linear precoding design in ELAA systems and motivate our investigation.

Different from current preconditioning techniques, we seek to regularize the extreme singular values of $\mathbf{A}$ with a rank-one regularization matrix. 
Through our investigation, we aim to answer the two questions:
{\em 1)} is it feasible to regularize single or multiple singular values with a rank-one matrix? {\em 2)} if yes, is there a low-complexity approach to improve $\kappa(\mathbf{A})$ for ELAA systems?

\section{Feasibility Analysis of Sherman-Morrison Regularization}\label{secIII}

In this section, we discuss the feasibility to regularize the extreme singular values of $\mathbf{A}$ with a rank-one matrix (denoted by $\mathbf{b}\mathbf{c}^H$, where $\mathbf{b}$, $\mathbf{c}\in\mathbb{C}^{N\times1}$) such that $\kappa(\mathbf{A}-\mathbf{b}\mathbf{c}^H)<\kappa(\mathbf{A})$.
After obtaining $(\mathbf{A}-\mathbf{b}\mathbf{c}^H)^{-1}$ through the HB algorithm, $\mathbf{A}^{-1}$ can be obtained through the Sherman-Morrison equation, which is reformulated as follows in \textit{Lemma \ref{lem1}} \cite{Petersen2012}.
This is the reason that the proposed approach is named Sherman-Morrison regularization.

\begin{lem}\label{lem1}
Given a matrix $\mathbf{A}$ and two vectors $\mathbf{b}$, $\mathbf{c}$, $\mathbf{A}^{-1}$ can be obtained from $(\mathbf{A}-\mathbf{b}\mathbf{c}^H)^{-1}$:
\begin{equation}
(\mathbf{A})^{-1}=(\mathbf{A}-\mathbf{b}\mathbf{c}^H)^{-1}-\frac{(\mathbf{A}-\mathbf{b}\mathbf{c}^H)^{-1}\mathbf{b}\mathbf{c}^H(\mathbf{A}-\mathbf{b}\mathbf{c}^H)^{-1}}{1+\mathbf{c}^H(\mathbf{A}-\mathbf{b}\mathbf{c}^H)^{-1}\mathbf{b}}.
\end{equation}
\end{lem}

With $\mathbf{X}_{t}\to(\mathbf{A}-\mathbf{b}\mathbf{c}^H)^{-1}$, $\mathbf{A}^{-1}$ 
is therefore be given by
\begin{equation}\label{eq14_1}
\mathbf{A}^{-1}=\mathbf{X}_t-\frac{\mathbf{X}_t\mathbf{b}\mathbf{c}^H\mathbf{X}_t}{1+\mathbf{c}^H\mathbf{X}_t\mathbf{b}}.
\end{equation}

It is perhaps worth noting that the Sherman-Morrison formula has been used in many scenarios for matrix inverse (e.g., quasi-Newton methods and matrix partitioning \cite{Hager1989})
However, to the best of our knowledge ,it has not been used to improve the condition number of a Wishart matrix.

\subsection{Regularization of the Largest Singular Value}\label{secIIIA}

When the SVD of $\mathbf{A}$ is available (remember $\mathbf{A}=\mathbf{U}\mathbf{\Lambda}\mathbf{U}^H$), it is feasible to change one specific singular value.
Let 
\begin{equation}\label{eq14}
\mathbf{b}=\mathbf{u}_0,~\mathbf{c}=\xi\mathbf{u}_0,
\end{equation}
where $\mathbf{u}_n$, $_{n=0,1,...,N-1}$ stands for the $n^\mathrm{th}$ column of $\mathbf{U}$ and $\xi\in\mathbb{R}$ for a scalar.
We have $\mathbf{b}\mathbf{c}^H$ to be a Hermitian matrix that shares the same unitary matrix as $\mathbf{A}$ in their SVDs.
Moreover, the largest singular value $\lambda_{0}$ in $\mathbf{A}$ becomes $|\lambda_0-\xi|$ in $(\mathbf{A}-\mathbf{b}\mathbf{c}^H)$, while other singular values remain the same, where $|\cdot|$ stands for the absolute value of a scalar.

\begin{thm}\label{thm1}
Given a Hermitian matrix $\mathbf{A}$ and a rank-one regularization matrix $\mathbf{b}\mathbf{c}^H$ given in \eqref{eq14}, the minimum condition number of $(\mathbf{A}-\mathbf{b}\mathbf{c}^H)$ is 
\begin{equation}
\kappa(\mathbf{A}-\mathbf{b}\mathbf{c}^H)_{\min}=(\lambda_{1})/(\lambda_{N-1}),
\end{equation}
where a sufficient condition to achieve this minimum is
\begin{equation}
\lambda_{0}-\lambda_{1}<\xi<\lambda_{0}-\lambda_{N-1}.
\end{equation}
\end{thm}
\begin{IEEEproof}
See Appendix \ref{appdxA}
\end{IEEEproof}

\textit{Theorem \ref{thm1}} reveals that with a proper $\xi$, the regularization matrix $\mathbf{b}\mathbf{c}^H$ can effectively cancel the largest singular value $\lambda_{0}$ to improve the condition number.
This would be particularly useful when the channel is highly correlated, e.g., when the channel is Rician and dominated by LoS paths (see Sec. \ref{secIVA} for further discussion).

\subsection{Regularization of Multiple Small Singular Values}
For small singular values, the problem is not exactly the same as in Sec. \ref{secIIIA}.
This is because in some MIMO systems (e.g., symmetric MIMO when the channel is i.i.d. Rayleigh fading) there are several extremely small singular values.
No doubt by letting $\mathbf{b}=\mathbf{u}_{N-1}$ and $\mathbf{c}=\xi\mathbf{u}_{N-1}$, the smallest singular value can be regularized to $\lambda_{N-1}-\xi$.
However, when there are several extremely small singular values, this could not make significant improvement.

This motivates us to investigate the feasibility to regularize multiple (i.e., $K$) small singular values at a time.
Different from \eqref{eq14}, here $\mathbf{b}$ and $\mathbf{c}$ are given by
\begin{equation}\label{eq18}
\mathbf{b}=\sum_{n=N-K}^{N-1}\mathbf{u}_{n},~\mathbf{c}=\xi\mathbf{b}.
\end{equation}

To analyze the singular values of $(\mathbf{A}-\mathbf{b}\mathbf{c}^H)$, we first identify the singular values that remain the same as in $\mathbf{A}$.
This can be done by having $\mathbf{U}^H(\mathbf{A}-\mathbf{b}\mathbf{c}^H)\mathbf{U}$ in \eqref{eq19}:
\begin{equation}\label{eq19}\footnotesize
\mathbf{U}^H(\mathbf{A}-\mathbf{b}\mathbf{c}^H)\mathbf{U}=\left[
\begin{matrix}
\lambda_0 & 0 & \cdots & 0 &  \\
0 & \lambda_1 & \ddots & \vdots &   \\
\vdots & \ddots & \ddots & 0 & \mathbf{0}^{(N-K)\times K}  \\
0 & \cdots & 0 & \lambda_{N-K} &    \\
 &  & \mathbf{0}^{K\times(N-K)} &  & \mathbf{\Psi}
\end{matrix}
\right],
\end{equation}
where $\mathbf{\Psi}\in\mathbb{R}^{K\times K}$ is a matrix whose non-diagonal entries are equal to $-\xi$:
\begin{equation}\label{eq20}\footnotesize
\mathbf{\Psi}=\left[
\begin{matrix}
\lambda_{N-K+1}-\xi & -\xi & \cdots & -\xi  \\
-\xi & \lambda_{N-K+2}-\xi & \ddots & \vdots  \\
\vdots & \ddots~~~~~~~~~~ & \ddots & -\xi  \\
-\xi &  \cdots & -\xi & \lambda_{N-1}-\xi
\end{matrix}
\right].
\end{equation}
The block-diagonal form of $\mathbf{U}^H(\mathbf{A}-\mathbf{b}\mathbf{c}^H)\mathbf{U}$ indicates that $\lambda_{0},...,\lambda_{N-K}$ are still singular values of $(\mathbf{A}-\mathbf{b}\mathbf{c}^H)$, and the rest of singular values (denoted by $\sigma_0,...,\sigma_{K}$) are determined by $\mathbf{\Psi}$.
Due to the space limit, here only the case when $K=2$ and $\xi>0$ is presented. In our transaction paper, we will present more complicated cases.

\begin{thm}\label{thm2}
Given the Hermitian matrix $\mathbf{\Psi}$ in \eqref{eq20} ($K=2$), a sufficient condition for its singular values ($\sigma_0$ and $\sigma_1$) to satisfy the following inequalities
\begin{equation}
\sigma_0>\lambda_{N-2},~\sigma_1>\lambda_{N-1}
\end{equation}
is that
\begin{equation}\label{eq22}
\xi>\frac{2\lambda_{N-2}(\lambda_{N-2}+\lambda_{N-1})}{3\lambda_{N-2}+\lambda_{N-1}}.
\end{equation}
\end{thm}
\begin{proof}
See Appendix \ref{appdxB}.
\end{proof}

Intuitively, $\mathbf{\Psi}$ may look like to converge to a rank-one matrix with the increase of $\xi$ where $\sigma_1\to0$. 
However, with some tidy-up work, the determinant of $\mathbf{\Psi}$ (denoted by $\det(\mathbf{\Psi})$) is actually given by $\det(\mathbf{\Psi})=\lambda_{N-2}\lambda_{N-1}-(\lambda_{N-2}+\lambda_{N-1})\xi$.
This shows that $\det(\mathbf{\Psi})\propto\xi$.
Since $\mathbf{\Psi}$ is Hermitian, $\sigma_0$ and $\sigma_1$ are also the eigenvalues of $\mathbf{\Psi}$. 
This implies that $\sigma_1$ does not tends to zero as $\xi$ tends to infinity.

Another issue is that $\sigma_0$ could be too big (i.e., $\sigma_0>\lambda_{0}$) and detrimental to $\kappa(\mathbf{A}-\mathbf{b}\mathbf{c}^H)$.
This case can be avoided with a properly set $\xi$.
Since $\sigma_0$ and $\sigma_1$ are also eigenvalues of $\mathbf{\Psi}$, we have $\|\mathbf{\Psi}\|_\mathrm{F}^2=\sigma_0^2+\sigma_1^2$, where $\|\mathbf{\Psi}\|_\mathrm{F}^2=4\xi^2-2(\lambda_{N-2}\lambda_{N-1})\xi+\lambda_{N-2}^2+\lambda_{N-1}^2$.
This means $\sigma_0^2$ is approximately upper bounded by $4\xi^2$.
Therefore, with a properly set $\xi$, $\sigma_0$ can be controlled to be no higher than $\lambda_{0}$.


\section{Low-Complexity Sherman-Morrison Regularization}\label{secIV}
Sec. \ref{secIII} showed the feasibility to regularize the extreme singular values through a rank-one regularization matrix designed based on the SVD of $\mathbf{A}$.
However, similar to the matrix inverse, the SVD has a complexity of $\mathcal{O}(N^3)$ and cannot be parallelized.
Therefore, it is not practical to assume the knowledge of SVD when designing $\mathbf{b}\mathbf{c}^H$.

In this section, we propose a low-complexity approach to regularize the largest or multiple small singular values based on the statistical property of the channel, respectively.

\subsection{LoS-Dominated ELAA}\label{secIVA}
In a ELAA system where the LoS paths dominate the antenna links (i.e., the channel is Rician with high K-factor), the channel is highly correlated (e.g., \cite{Wang2022b}). 
In this case, the entries of $\mathbf{A}$ are nearly equal to each other, and there is a dominant singular value $\lambda_{0}$: $\lambda_{0}\gg\lambda_{1}>...>\lambda_{N-1}$.
This is a suitable scenario for \textit{Theorem \ref{thm1}}, where regularizing the largest singular value would significantly improve $\kappa(\mathbf{A})$.

Since $\lambda_{0}$ is so dominant, $\mathbf{A}$ is nearly a rank-one matrix where $\mathbf{b}=\mathbf{u}_0$ (as given in \eqref{eq14}) can be approximately obtained from any column of $\mathbf{A}$: $\mathbf{b}\approx(\mathbf{a}_n)/|\mathbf{a}_n|$.
Note that here $\mathbf{b}$ is normalized to fit the norm of $\mathbf{u}_0$.

However, if we assume $\mathbf{c}$ to be a scaled vector of $\mathbf{b}$, it is hard to decide the value of $\xi$, who approximately indicates the portion of $\lambda_{0}$ among the singular values.
Alternatively, we may seek to know the portion of $\mathbf{b}$ in each column of $\mathbf{A}$ to know $\xi$, and this is done by having $\mathbf{c}=\mathbf{A}\mathbf{b}$.
But this alternative approach would still be too naive, as the rank of $(\mathbf{A}-\mathbf{b}\mathbf{c}^H)$ will reduce to $N-1$ in this case, and $(\mathbf{A}-\mathbf{b}\mathbf{c}^H)$ would have infinite condition number.

To make the alternative approach effective, we propose to deduct a scaled identity matrix from $\mathbf{A}$ to help obtaining $\mathbf{c}$ as follows.
\begin{equation}\label{eq26}
\mathbf{\Delta}\triangleq\mathbf{A}-\alpha\mathbf{I}.
\end{equation}
Since $\mathbf{A}$ is Hermitian, $\mathbf{\Delta}$ shares the same unitary matrix of SVD with $\mathbf{A}$, and the singular values of $\mathbf{\Delta}$ becomes $\lambda_{n}-\alpha$.
When $\alpha$ is small (i.e., $\alpha\ll\lambda_{0}$), $\lambda_{0}-\alpha$ is still the dominant singular value of $\mathbf{\Delta}$, and it is reasonable to obtain $\mathbf{b}$ and $\mathbf{c}$ from $\mathbf{\Delta}$:
\begin{equation}\label{eq27}
\mathbf{b}=\boldsymbol{\delta}_n/|\boldsymbol{\delta}_n|,~\mathbf{c}=\mathbf{\Delta}\mathbf{b}.
\end{equation}

To determine $\alpha$, it is assumed that $\|\mathbf{A}\|_\mathrm{F}^2$ is known.
Since $\mathbf{A}$ is Hermitian, we have $\|\mathbf{A}\|_\mathrm{F}^2=\sum_{n=0}^{N-1}\lambda_{n}^2$, where $\lambda_{0}$ is dominant.
To ensure $\alpha\ll\lambda_{0}$, it is reasonable to assume $\alpha$ is smaller than the average of singular values: $\alpha\ll\|\mathbf{A}\|_\mathrm{F}^2/(N)$.
These discussions are summarized in \textit{Remark \ref{rmk1}}
\begin{rmk}\label{rmk1}
In a LoS-dominated ELAA system where $\mathbf{A}$ has a dominant singular value, the regularization matrix to improve $\kappa(\mathbf{A})$ is approximately given by \eqref{eq27}, and the scalar $\alpha$ to obtain $\mathbf{\Delta}$ satisfies that $\alpha\ll\|\mathbf{A}\|_\mathrm{F}^2/(N)$.
\end{rmk}

{It is observed from \textit{Remark \ref{rmk1}} that obtaining the rank-one regularization matrix involves only vector-matrix multiplication, and the complexity is therefore $\mathcal{O}(N^2)$ \cite{8417575}.}

\subsection{Large Symmetric ELAA}
Although LoS paths are an important feature of ELAA systems, it is no longer the dominant issue with the increase of MIMO size.
And the ever-growing demand of spectrum efficiency is pushing the multi-user MIMO back to the symmetric architecture \cite{Wang2022c}.
In this case, the system is no longer underdetermined as conventional massive-MIMO.
This alone makes the channel very ill-conditioned.

Since the analysis of the statistical behavior in a symmetric ELAA system is still lacking in the literature, here we use the statistical behavior of i.i.d. Rayleigh fading symmetric MIMO as alternative (i.e., $\mathbf{H}\sim\mathcal{CN}(0,\mathbf{I})$), where the ill-condition of $\mathbf{A}$ is caused by several extremely small singular values \cite{Wang2022c}.
This is a suitable scenario for \textit{Theorem \ref{thm2}}.
In the simulation results, we will show that our low-complexity approach is also effective for symmetric ELAA systems.

Compared to the LoS-dominated case, it is hard to obtain $\mathbf{b}$ that mainly consists of $\mathbf{u}_{N-2}$ and $\mathbf{u}_{N-1}$ from $\mathbf{A}$.
To handle this challenge, we first need to understand that when $\xi\ll\lambda_{0}$, $\xi$ can still satisfy the inequality of \textit{Theorem \ref{thm2}}. 
This is because the right side of \eqref{eq22} satisfies
\begin{equation}\label{eq28}
\frac{2\lambda_{N-2}(\lambda_{N-2}+\lambda_{N-1})}{3\lambda_{N-2}+\lambda_{N-1}}<\frac{4\lambda_{N-2}^2}{3\lambda_{N-2}}=\frac{4\lambda_{N-2}}{3}.
\end{equation}
\eqref{eq28} shows that even when $\xi\approx\lambda_{N-2}$, \textit{Theorem \ref{thm2}} can still be satisfied.
In a large symmetric MIMO, $\lambda_{N-2}$ is often lower than $10^{-3}$ (i.e., $\lambda_{N-2}\ll\lambda_{0}$), and therefore the $\xi$ can satisfy \textit{Theorem \ref{thm2}} when $\xi\ll\lambda_{0}$.
Therefore, even when $\mathbf{b}$ contains singular vectors of high singular values (e.g., $\mathbf{u}_0$ and $\mathbf{u}_1$), the non-diagonal elements in $\mathbf{U}^H\mathbf{A}\mathbf{U}$ are negligible compared to its diagonals, and the change of large singular values in $\mathbf{A}$ is negligible.
Hence, we can focus on the analysis of small singular values when $\xi$ is small.

Another challenge is that since $\lambda_{N-2}$ and $\lambda_{N-1}$ are so small, any column of $\mathbf{A}$ hardly contains $\mathbf{u}_{N-1}$ or $\mathbf{u}_{N-1}$.
To handle this challenge, we propose to raise the value of $\alpha$ ($\alpha>0$) in \eqref{eq26}.
The reason is that the singular values of $\mathbf{\Delta}$ is $|\lambda_{n}-\alpha|$.
With $\alpha\gg\lambda_{N-2}$, $|\lambda_{N-2}-\alpha|$ and $|\lambda_{N-1}-\alpha|$ are no longer the smallest singular value of $\mathbf{\Delta}$.
By letting $\mathbf{b}=\boldsymbol{\delta}_n/|\boldsymbol{\delta}_n|$, we are able to obtain $\mathbf{b}$ that contains $\mathbf{u}_{N-2}$ and $\mathbf{u}_{N-1}$.

Similar to the case in Sec. \ref{secIVA}, we approximate $\mathbf{c}=\xi\mathbf{b}$ to be $\mathbf{\Delta}\mathbf{b}$ as in \eqref{eq27}.
In this case, $\xi$ is approximated as $\|\mathbf{c}\|^2$.
Since $\|\mathbf{c}\|^2=\|\mathbf{\Delta}\mathbf{b}\|^2\leq\|\mathbf{\Delta}\|^2\|\mathbf{b}\|^2=\|\mathbf{\Delta}\|^2$, we can ensure $\xi\ll\lambda_{0}$ by limiting $\|\mathbf{\Delta}\|^2$.
For a large symmetric MIMO, we have $a_{n,n}\to1$ with the increase of $N$.
Hence, we can set $\alpha=1$ to reduce $\|\mathbf{\Delta}\|^2$.
These discussions are summarized in \textit{Remark \ref{rmk2}}.

\begin{rmk}\label{rmk2}
In a large symmetric MIMO where the channel is i.i.d. Rayleigh fading, the regularization matrix to improve $\kappa(\mathbf{A})$ is approximately given by \eqref{eq27} with the scalar $\alpha=1$.
\end{rmk}

Overall, \textit{Remark \ref{rmk1}} and \textit{Remark \ref{rmk2}} show that without the knowledge of SVD, the Sherman-Morrison regularization can still be implemented with low-complexity in a general mathematical form.
For different ELAA systems, we only need to change $\alpha$ based on the statistical property of the channel.
{Similar to \textit{Remark \ref{rmk1}}, the complexity of \textit{Remark \ref{rmk2}} is still $\mathcal{O}(N^2)$.}

\subsection{Combination with the List Algorithm}\label{secIVC}
{the complexity of the Sherman-Morrison regularization is }
Although \textit{Remark \ref{rmk2}} provides a low-complexity regularization approach, it cannot guarantee that $\mathbf{b}=\boldsymbol{\delta}_n/|\boldsymbol{\delta}_n|$ consists of $\mathbf{u}_{N-2}$ and $\mathbf{u}_{N-1}$ for arbitrary $n$.
To handle this issue, we propose to combine the list algorithm with the low-complexity approach.
Denote $\mathbf{b}=\boldsymbol{\delta}_n/|\boldsymbol{\delta}_n|$ by $\mathbf{b}_n$ and its corresponding $\mathbf{c}$ by $\mathbf{c}_n$.
The idea is to calculate $(\mathbf{A}-\mathbf{b}_n\mathbf{c}_n^H)^{-1}$,  $_{n=0,...N-1}$ using the HB algorithm in parallel.
Then, we choose the best candidate such that 
\begin{equation}
\{\mathbf{b}^\star,\mathbf{c}^\star\}=\argmin_{n}\|\mathbf{I}-(\mathbf{A}-\mathbf{b}_n\mathbf{c}_n^H)\mathbf{X}_t\|_\mathrm{F}^2.
\end{equation}
Finally, $\mathbf{A}^{-1}$ is obtained based on $(\mathbf{A}-\mathbf{b}^\star\mathbf{c}^{\star H})^{-1}$ through \textit{Lemma \ref{lem1}}.
It is anticipated that the list algorithm could greatly improve the condition number in symmetric ELAA systems.

\section{Simulation Results and Discussions}\label{secV}
In this section, computer simulations are carried out using MATLAB to demonstrate the performance of the proposed low-complexity Sherman-Morrison regularization in terms of average {symbol error rate (SER)} and iteration number. {The channel-coded results will be shown in our future transaction paper.}
The proposed regularization approach is compared to current preconditioning techniques. 
Denote the diagonal of $\mathbf{A}$ by $\mathbf{D}$ and the strict lower part of $\mathbf{A}$ by $\mathbf{L}$, the preconditioner $\mathbf{P}$ for the Jacobi, GS and SSOR methods are given by $\mathbf{D}^{-1}$, $(\mathbf{D}+\mathbf{L})^{-1}$ and $(\mathbf{L}^H+\mathbf{D})^{-1}\mathbf{D}^{-1}(\mathbf{L}+\mathbf{D})^{-1}$, respectively.
{Each approach is only different in initialization, and experiences the same iterative process of HB algorithm. Hence, the iteration number is used to compare the convergence speed.}

In the simulation, it is not suitable to assume $\omega^\star$ is available, since it requires the knowledge of singular values of $\mathbf{A}$.
Alternatively, the Gershgorin circle theorem is adopted to ensure the convergence of the HB algorithm \cite{Saad2003}.
Taking the example when $\mathbf{X}_0=\omega\mathbf{A}$, $\omega$ determined by the Gershgorin circle theorem is given by $\omega_\mathrm{Ger}=\big(\max_n(\sum_{i=0}^{N-1}|\mathbf{a}_{n}^H\mathbf{a}_{i}|)\big)^{-1}$.

The carrier frequency is set to be $3.5$ GHz and the modulation is $64$ QAM \cite{Wang2022c}.
For ELAA systems, the channel model in \cite{9685536} is employed in appreciation of the comprehensive consideration of path loss, shadow fading as well as LoS states.
Specifically in our simulation, the height of users is $1.5$ m, the height of the base station is $10$ m, and the distance between adjacent receiver antennas in each user is $0.0429$ m (i.e., half wavelength of the carrier frequency).
During the simulation, the channel is normalized such that $\|\mathbf{H}\|_\mathrm{F}^2=N$ to remove the power gain brought by the path loss of LoS antenna links.

{\em Experiment $1$:}
The aim of this experiment is to investigate the performance of Sherman-Morrison regularization in asymmetric LoS-dominated ELAA systems.
Specifically, we consider $M=128$ and $N=16$ ($2$ users with $8$ antennas each user) and there are only LoS antenna links.

\begin{figure}[t]
\centering
\includegraphics[scale=0.47]{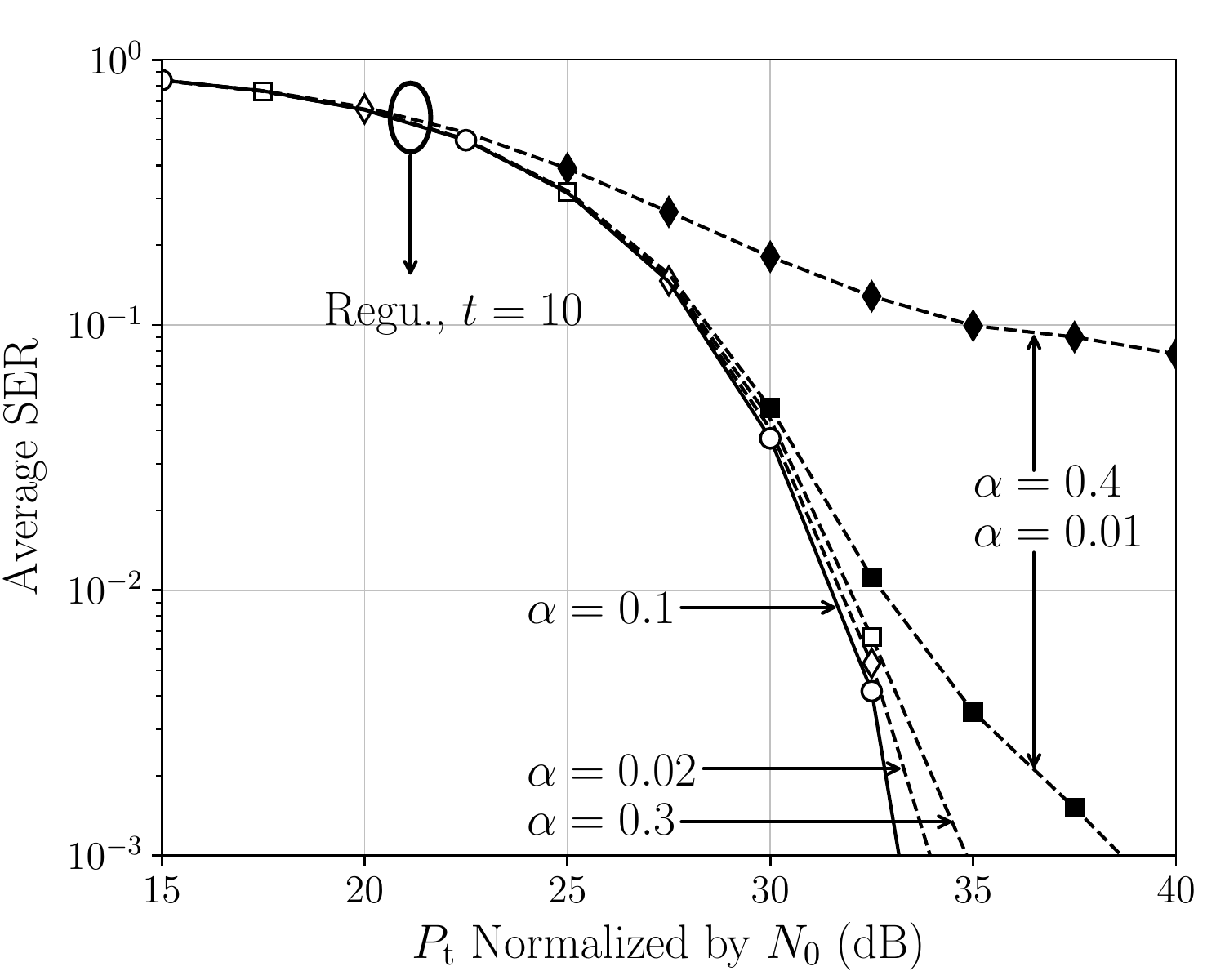}
\caption{Average SER of the Sherman-Morrison regularization for ZF with the change of $\alpha$ when $M=128$, $N=16$ in LoS-dominated ELAA systems.}
\vspace{-1.5em}
\label{fig1}
\end{figure}

\begin{figure}[t]
\centering
\includegraphics[scale=0.47]{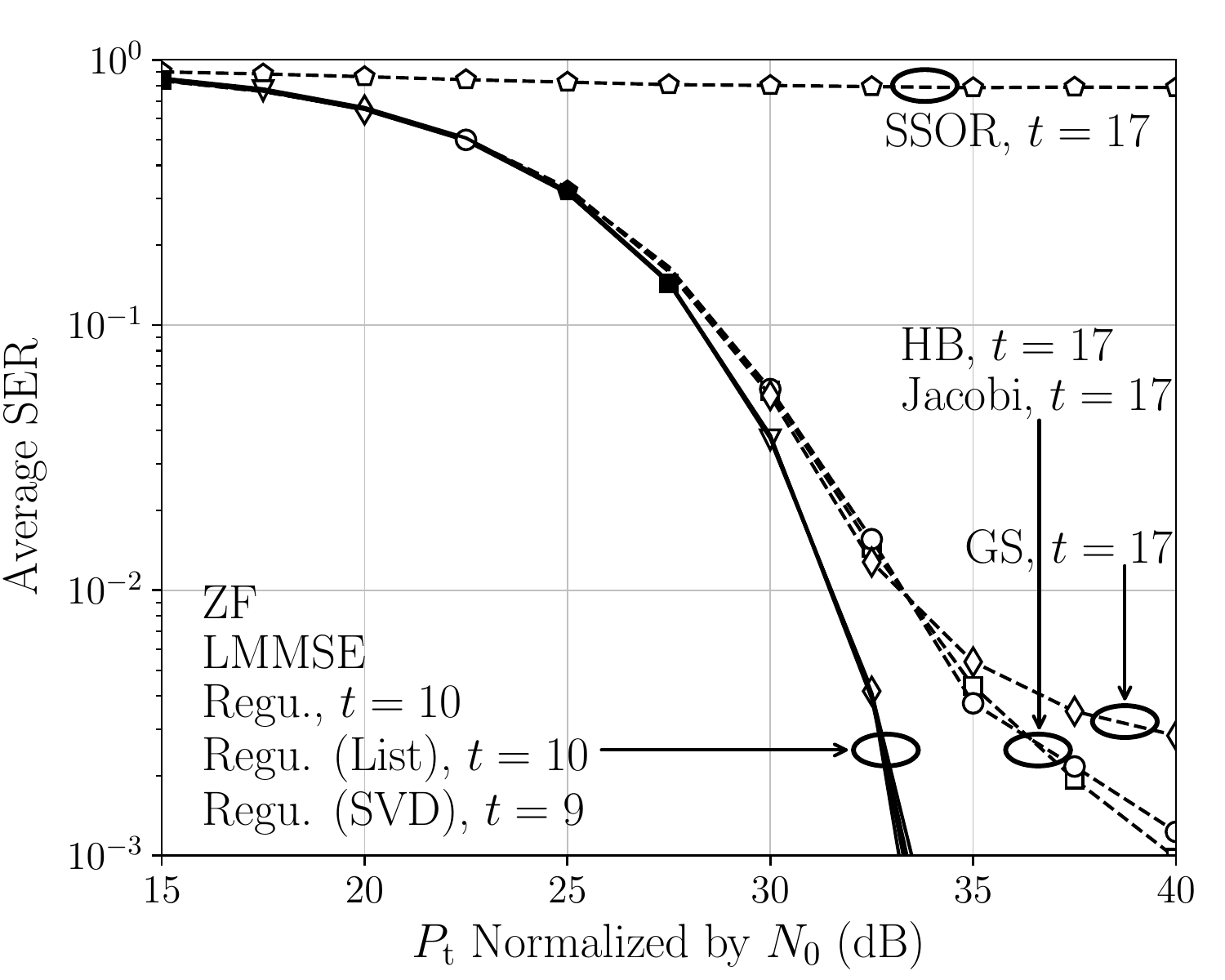}
\caption{Average SER of the Sherman-Morrison regularization for ZF compared to preconditioning when $M=128$, $N=16$, $\alpha=0.1$.}
\vspace{-1.5em}
\label{fig2}
\end{figure}

Fig. \ref{fig1} shows the average SER of the Sherman-Morrison regularization with respect to the change of $\alpha$ when the iteration number is $10$ for ZF precoding.
It can be observed that the regularization approach achieves the best performance when $\alpha=0.1$ (i.e., $\alpha\ll1$). 
This coincides to the discussions in \textit{Remark \ref{rmk1}}.
When $\alpha$ is too small, the performance of the regularization is degraded.
This is because in that case, $\alpha$ is almost zero, and the rank of $(\mathbf{A}-\mathbf{b}\mathbf{c}^H)$ is nearly $(N-1)$.
In Fig, \ref{fig2}, we will use the case of $\alpha=0.1$ to compare the performance of the regularization approach to preconditioning.

Fig. \ref{fig2} shows the average SER of the Sherman-Morrison regularization (solid lines) compared to preconditioning or using the HB algorithm alone (dash lines). 
The iteration number $10$ and $17$ for the regularization approach and other approaches, respectively.
The first thing to be noticed is that with only $10$ iterations, the regularization approach has already converged to the performance of ZF and LMMSE.
Compared to the HB algorithm who has a $6$ dB performance degradation from ZF, the regularization can reduce at least $7$ iterations (i.e., more than $40\%$ of reduction).
Secondly, it can be observed that using preconditioning could hardly bring improvement.
This is because the entries of $\mathbf{A}$ are all nearly equal, which breaks the assumption that $\mathbf{A}$ is diagonal dominant.

{\em Experiment $2$:}
The aim of this experiment is to investigate the performance of Sherman-Morrison regularization in symmetric MIMO both for i.i.d. Rayleigh fading channel and ELAA channel.
We consider $M=128$ and $N=128$ ($16$ users with $8$ antennas each user) as well as mixed LoS/non-LoS antenna links.
As discussed in \textit{Remark \ref{rmk2}}, $\alpha=1$.

\begin{figure}[t]
\centering
\includegraphics[scale=0.47]{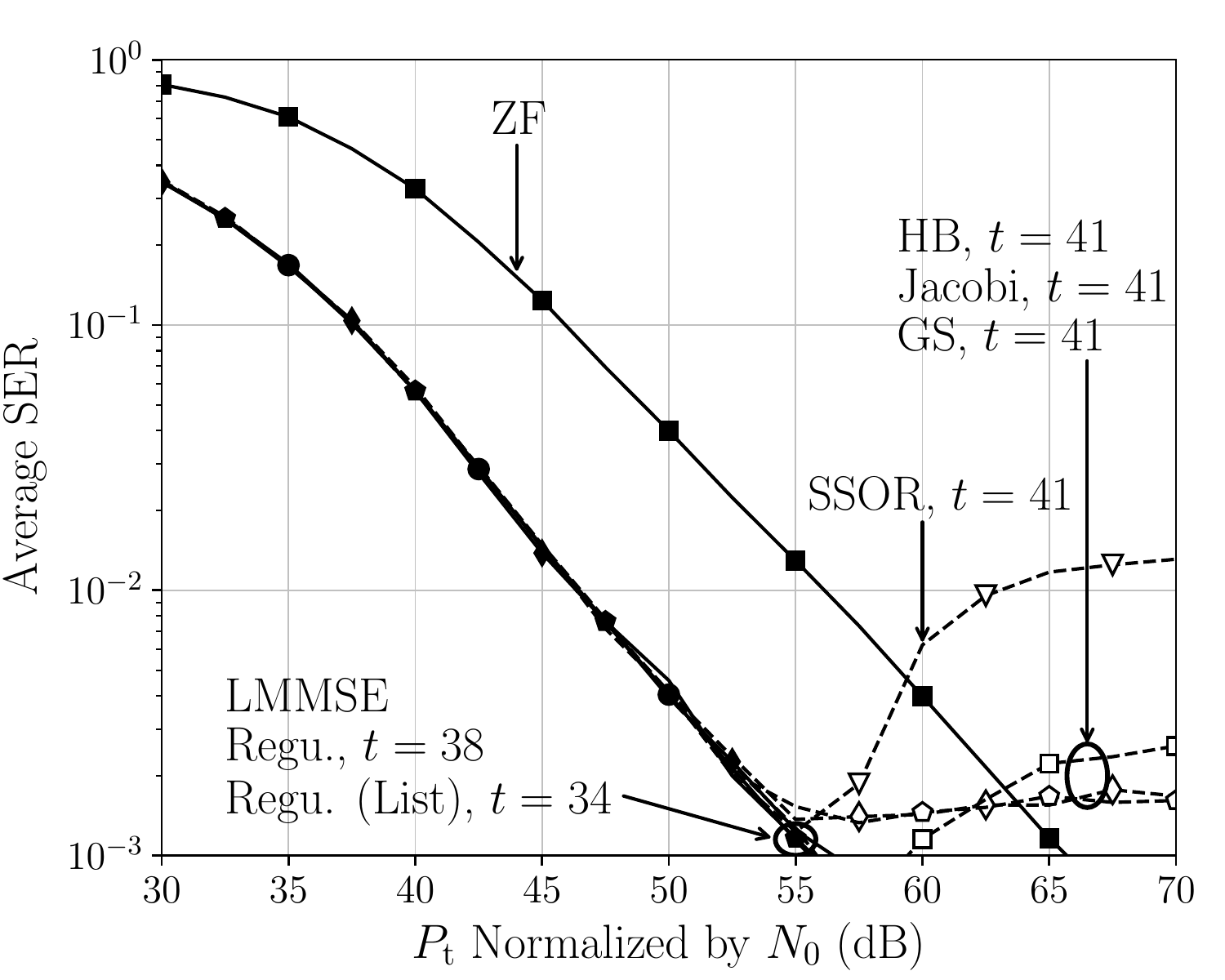}
\caption{Average SER of the Sherman-Morrison regularization for LMMSE compared to preconditioning when $M=128$, $N=128$ (i.i.d. Rayleigh).}
\vspace{-1.5em}
\label{fig3}
\end{figure}

Fig. \ref{fig3} shows the average SER of the Sherman-Morrison regularization (solid lines) compared to preconditioning or using the HB algorithm alone (dash lines) for LMMSE precoding.
The iteration number is $38$ and $34$ for the regularization approach with or without list algorithm, respectively.
For the baselines, the iteration number is $41$.
It is observed that the GS and SSOR indeed converge to LMMSE performance before $P_\mathrm{t}=55$ dB, but their SER raises up as $P_\mathrm{t}$ increases.
This is because the term $N_0/P_\mathrm{t}\mathbf{I}$ of LMMSE is enhancing the diagonal of $\mathbf{A}$, which is beneficial for GS and SSOR.
The Jacobi preconditioning has similar performance as the HB algorithm, where both of them reaches an error floor of around $2\times10^{-3}$ (lower than GS and SSOR).
Overall, the preconditioning can hardly improve the performance compared to using the HB algorithm alone.

In comparison, the regularization approach fully converges to LMMSE performance.
The iteration number is reduced for $3$ and $7$ (with list algorithm) compared to using the HB algorithm alone (i.e., around $7\%$ and $17\%$ reduction).
This shows that the regularization approach significantly outperforms the preconditioning techniques in symmetric MIMO.
As anticipated in Sec. \ref{secIVC}, the list algorithm can greatly benefit the performance.

\begin{figure}[t]
\centering
\includegraphics[scale=0.47]{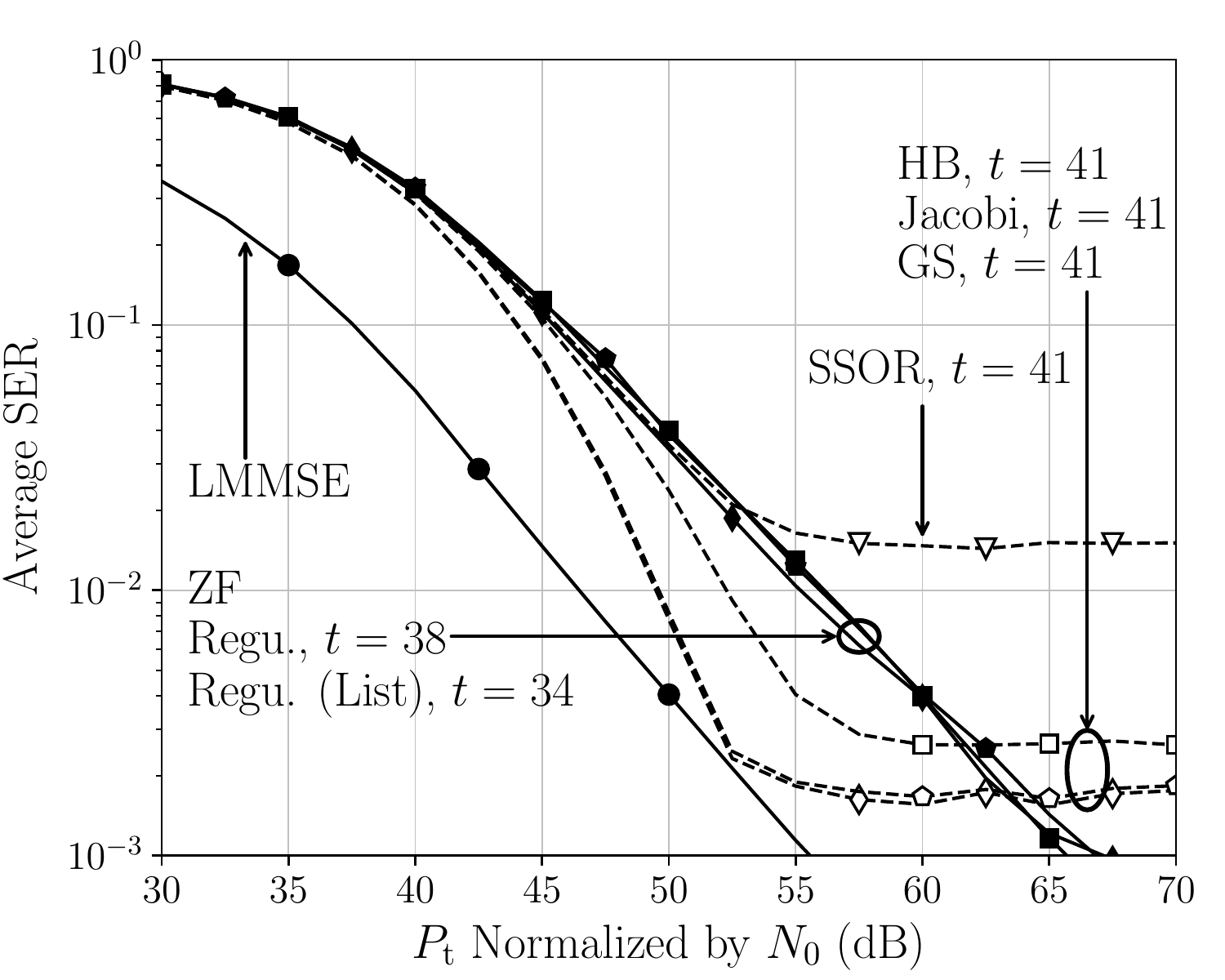}
\caption{Average SER of the Sherman-Morrison regularization for ZF compared to preconditioning when $M=128$, $N=128$ (i.i.d. Rayleigh).}
\vspace{-1.0em}
\label{fig4}
\end{figure}

Fig. \ref{fig4} shows the average SER of the regularization approach for ZF precoding.
With the same iteration number as in Fig. \ref{fig3}, the regularization approach achieves close performance to ZF precoding (within $1$ dB).
While for the baselines, the interesting phenomenon is that the GS, Jacobi and HB algorithm outperforms ZF precoding when $P_\mathrm{t}<63$ dB.
This is because the error in $\mathbf{X}_t$ has similar effect as the term $N_0/P_\mathrm{t}\mathbf{I}$ in LMMSE.
Nevertheless, this also results in an error floor.
Moreover, when ZF precoding is combined with other techniques (e.g., nonlinear precoding \cite{Wang2022c}), a more accurate knowledge of $\mathbf{A}^{-1}$ is needed, where the regularization approach is in favor.

\begin{figure}[t]
\centering
\includegraphics[scale=0.47]{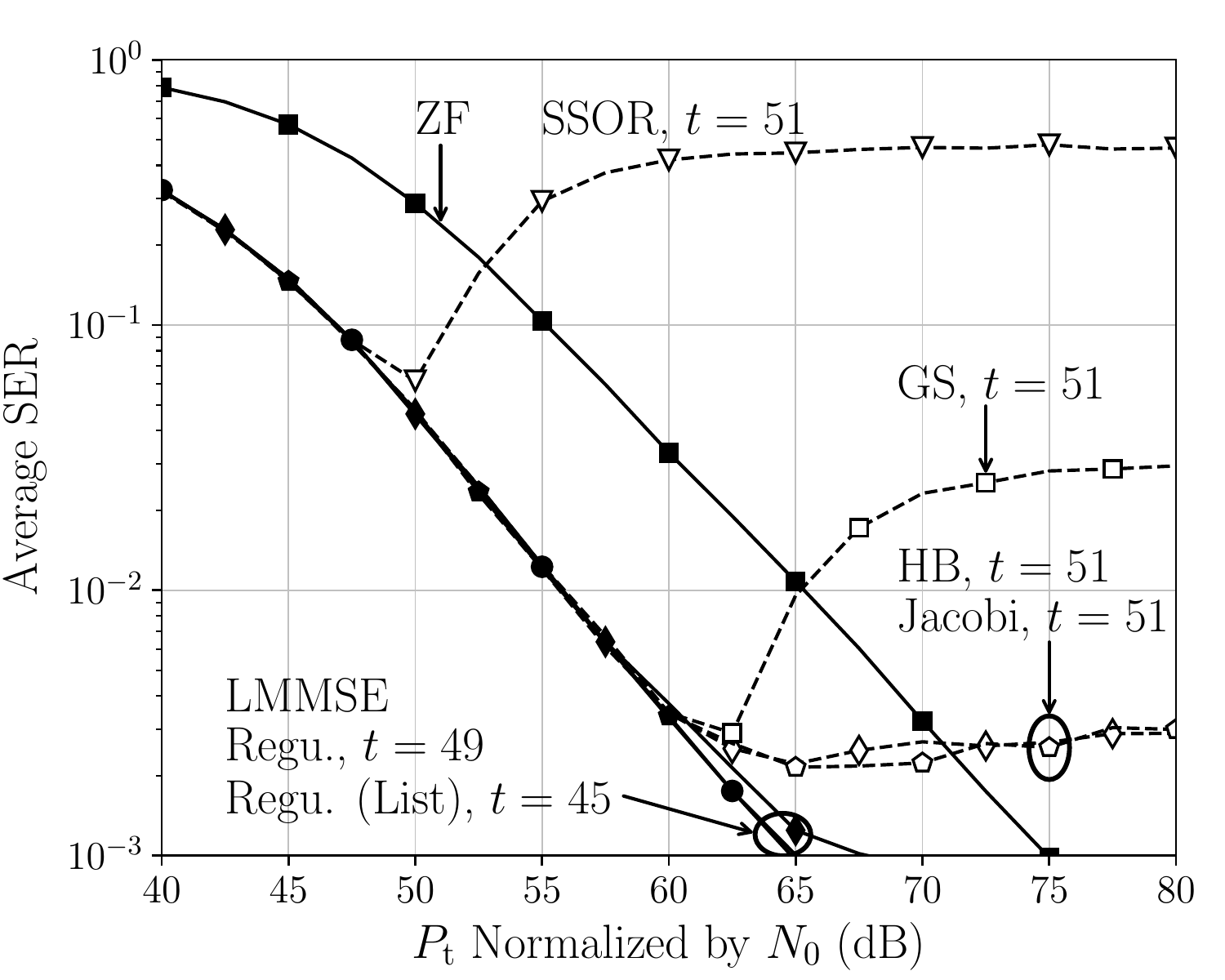}
\caption{Average SER of the Sherman-Morrison regularization for LMMSE compared to preconditioning when $M=128$, $N=128$ (ELAA).}
\vspace{-1.5em}
\label{fig5}
\end{figure}

Fig. \ref{fig5} shows the average SER of the regularization approach for LMMSE precoding in an ELAA system.
Overall, the spatial non-stationarity brings the performance degradation, and the GS and SSOR are suffering more from the raise-back of SER.
However, the regularization approach can still achieve close performance to LMMSE with smaller iteration number ($2$ and $6$ (with list algorithm), respectively).
This shows the regularization approach remain effective in symmetric ELAA systems, while the preconditioning techniques can hardly accelerate the convergence.

\section{Conclusion}\label{secVI}
In this paper, it has been proven that it is feasible to regularize the extreme singular values of a matrix with a rank-one regularization matrix.
In light of the feasibility analysis, a low-complexity approach, namely the Sherman-Morrison regularization, has been proposed to approximate the regularization matrix based on the statistical property of the channel so as to improve the matrix condition.
Computer simulations have demonstrated that the proposed regularization approach can significantly reduce the iteration number required achieve the convergence of iterative linear precoding, while current preconditioning techniques can hardly bring any improvement.
In LoS-dominated ELAA systems, the proposed approach can reduce the iteration number for $40\%$. 
In symmetric ELAA systems, the proposed approach can still reduce the iteration number for $10\%$ with the help of list algorithm.

\appendices
\section{Proof of Theorem \ref{thm1}}\label{appdxA}
\begin{IEEEproof}
We first prove that the minimum of $\kappa(\mathbf{A}-\mathbf{b}\mathbf{c}^H)$ is $(\lambda_{1})/(\lambda_{N-1})$.
The only changed singular value is that $\lambda_{0}$ becomes $|\lambda_{0}-\xi|$. When $|\lambda_{0}-\xi|<\lambda_{N-1}$, $\kappa(\mathbf{A}-\mathbf{b}\mathbf{c}^H)=(\lambda_{1})/|\lambda_{0}-\xi|>(\lambda_{1})/(\lambda_{N-1})$.
When $|\lambda_{0}-\xi|>\lambda_{1}$, $\kappa(\mathbf{A}-\mathbf{b}\mathbf{c}^H)=|\lambda_{0}-\xi|/(\lambda_{N-1})>(\lambda_{1})/(\lambda_{N-1})$.
Only when $\lambda_{N-1}<|\lambda_{0}-\xi|<\lambda_{1}$, $\kappa(\mathbf{A}-\mathbf{b}\mathbf{c}^H)=(\lambda_{1})/(\lambda_{N-1})$, which is the minimum value.

When $\lambda_{0}-\lambda_{1}<\xi<\lambda_{0}-\lambda_{N-1}$, $(\lambda_{0}-\xi)$ satisfies that
\begin{equation}\label{apddxEq01}
0<\lambda_{N-1}<\lambda_{0}-\xi<\lambda_{0}.
\end{equation}
Hence, we have $\lambda_{N-1}<|\lambda_{0}-\xi|<\lambda_{1}$, and the minimum of $\kappa(\mathbf{A}-\mathbf{b}\mathbf{c}^H)$ is achieved.
\end{IEEEproof}

\section{Proof of Theorem \ref{thm2}}\label{appdxB}
\begin{IEEEproof}
Based on the expression of $\mathbf{\Psi}$, $\sigma_0$ and $\sigma_1$ can be given by
\begin{equation}\label{apddxEq02}\small
\sigma_0=\frac{\sqrt{L_1+\sqrt{L_1^2-4L_2}}}{\sqrt{2}},~\sigma_1=\frac{\sqrt{L_1-\sqrt{L_1^2-4L_2}}}{\sqrt{2}},
\end{equation}
where $L_1$, $L_2$ are given by
\begin{IEEEeqnarray}{rl}
\small
L_1 &= \lambda_{N-2}^2+\lambda_{N-1}^2-2\lambda_{N-2}\xi-2\lambda_{N-1}\xi+4\xi^2,\label{apddxEq03}\\
L_2 &= \lambda_{N-2}^2\lambda_{N-1}^2-2\lambda_{N-2}^2\lambda_{N-1}\xi-2\lambda_{N-2}\lambda_{N-1}^2\xi\IEEEnonumber\\
&~~~~~~~~~~+\lambda_{N-2}^2\xi^2+2\lambda_{N-2}\lambda_{N-1}\xi^2+\lambda_{N-1}^2\xi^2.\label{apddxEq04}
\end{IEEEeqnarray}

{\em 1) $\sigma_0>\lambda_{N-2}$:} 
Since $\sigma_0>0$ and $\lambda_{N-2}>0$ (as they are singular values), this inequality means
\begin{equation}\label{apddxEq05}
\sigma_0^2>\lambda_{N-2}^2.
\end{equation} 
Substitute \eqref{apddxEq02} into \eqref{apddxEq05} and after some tidy-up work, we have
\begin{equation}\label{apddxEq06}
L_1-2\lambda_{N-2}^2>-\sqrt{L_1^2-4L_2}.
\end{equation}
When $L_1-2\lambda_{N-2}^2>0$, it is clear that \eqref{apddxEq06} holds since the right side is negative.
When $L_1-2\lambda_{N-2}^2<0$, \eqref{apddxEq06} means 
\begin{equation}\label{apddxEq07}
(L_1-2\lambda_{N-2}^2)^2<L_1^2-4L_2.
\end{equation}
\eqref{apddxEq07} can be simplified as
\begin{equation}\label{apddxEq08}
\lambda_{N-2}^4<L_1\lambda_{N-2}^2-L_2.
\end{equation}
Substitute \eqref{apddxEq03} and \eqref{apddxEq04} into \eqref{apddxEq08}, we have
\begin{IEEEeqnarray}{rl}
\lambda_{N-2}^4&<\lambda_{N-2}^4-2\lambda_{N-2}^3\xi+2\lambda_{N-2}\lambda_{N-1}^2\xi\IEEEnonumber\\
&~~-2\lambda_{N-2}\lambda_{N-1}\xi^2+3\lambda_{N-2}^2\xi^2-\lambda_{N-1}\xi^2.\label{apddxEq09}
\end{IEEEeqnarray}
It can be seen that $\lambda_{N-2}^4$ on both sides can be canceled. 
For the rest terms, $\xi$ can be canceled.
Since $\xi>0$,  \eqref{apddxEq09} can be simplified as:
\begin{equation}\label{apddxEq10}
2\lambda_{N-2}^2-2\lambda_{N-2}\lambda_{N-1}^2<(3\lambda_{N-2}^2-2\lambda_{N-2}\lambda_{N-1}-\lambda_{N-1}^2)\xi.
\end{equation}
After factorization, it can be known that both side of \eqref{apddxEq10} contains the term $\lambda_{N-2}-\lambda_{N-1}$. 
Since $\lambda_{N-2}-\lambda_{N-1}>0$, it can be canceled such that 
\begin{equation}\label{apddxEq11}\small
\xi>\frac{2\lambda_{N-2}(\lambda_{N-2}+\lambda_{N-1})}{3\lambda_{N-2}+\lambda_{N-1}}.
\end{equation}

\eqref{apddxEq11} is a sufficient condition for $\sigma_0>\lambda_{N-2}$ (denote the right side of \eqref{apddxEq11} by $\xi_\mathrm{T1}$ for notation simplicity).
If $L_1-2\lambda_{N-2}^2<0$ when $\xi=\xi_\mathrm{T1}$, $\xi_\mathrm{T1}$ shows the lowest value of $\xi$ to satisfy \eqref{apddxEq05}.
Otherwise, if $L_1-2\lambda_{N-2}^2>0$ when $\xi=\xi_\mathrm{T1}$, $L_1-2\lambda_{N-2}^2$ will increase with $\xi$ when $\xi>\xi_\mathrm{T1}$.
This is because $L_1-2\lambda_{N-2}^2=0$ is a quadratic equation of $\xi$ and has only one positive root.
Moreover, $L_1-2\lambda_{N-2}^2$ has positive quadratic terms of $\xi^2$.

{\em 2) $\sigma_1>\lambda_{N-1}$:} 
$\sigma_1>\lambda_{N-1}$ means
\begin{equation}\label{apddxEq12}
\sigma_1^2>\lambda_{N-1}^2,
\end{equation}
which is
\begin{equation}\label{apddxEq13}
L_1-2\lambda_{N-1}^2>\sqrt{L_1^2-4L_2}.
\end{equation}
\eqref{apddxEq13} indicates that both side are positive, and is equivalent to
\begin{equation}\label{apddxEq14}
L_1^2-4L_1\lambda_{N-1}^2+4\lambda_{N-1}^4>L_1^2-4L_2.
\end{equation}
\eqref{apddxEq14} can be simplified as
\begin{equation}\label{apddxEq15}
\lambda_{N-1}^4>L_1\lambda_{N-1}^2-L_2.
\end{equation}
Substitute \eqref{apddxEq03} and \eqref{apddxEq04} into \eqref{apddxEq15}, we have
\begin{IEEEeqnarray}{rl}\small
\lambda_{N-1}^4&>\lambda_{N-1}^4-2\lambda_{N-1}^3\xi+2\lambda_{N-2}^2\lambda_{N-1}\xi\IEEEnonumber\\
&~~-2\lambda_{N-2}\lambda_{N-1}\xi^2-\lambda_{N-2}^2\xi^2+3\lambda_{N-1}\xi^2.\label{apddxEq16}
\end{IEEEeqnarray}
With similar simplification procedure as \eqref{apddxEq10}, \eqref{apddxEq16} leads to
\begin{equation}\label{apddxEq17}
\xi>\frac{2(\lambda_{N-2}+\lambda_{N-1})\lambda_{N-1}}{\lambda_{N-2}+3\lambda_{N-1}}.
\end{equation}
Denote the right side of \eqref{apddxEq17} by $\xi_\mathrm{T2}$.
Then, we compare $\xi_{\mathrm{T}1}$ and $\xi_{\mathrm{T}2}$:
\begin{equation}\small
\frac{\xi_{\mathrm{T}1}}{\xi_{\mathrm{T}2}}=\frac{\lambda_{N-2}^2+3\lambda_{N-2}\lambda_{N-1}}{\lambda_{N-1}^2+3\lambda_{N-2}\lambda_{N-1}}>1.
\end{equation}

Hence, when $\xi>\xi_{\mathrm{T}1}$, we have $\sigma_1>\lambda_{N-1}$ as well. 
\textit{Theorem \ref{thm2}} is proved.
\end{IEEEproof}

\section*{Acknowledgement}
This work was funded by the 5G Innovation Centre and the 6G Innovation Centre.

\balance

\ifCLASSOPTIONcaptionsoff
\newpage
\fi

\bibliographystyle{myIEEEtran}
\bibliography{./../URLLC,./../Bib_Else,./../Books_and_Standards,./../NLP_Downlink,./../GroupPaper,./../IterativePrecoding}		
\end{document}